\documentclass{article}
\usepackage{fullpage}
\usepackage{amsmath}
\usepackage{txfonts,natbib}
\usepackage{graphicx,color}

\definecolor{gray}{rgb}{0.5,0.5,0.5}

\newcommand{\bvec}{\mathbf}

\DeclareRobustCommand{\ion}[2]{%
\relax\ifmmode
\ifx\testbx\f@series
{\mathbf{#1\,\textsc{#2}}}\else
{\mathrm{#1\,\textsc{#2}}}\fi
\else\textup{#1\,{\mdseries\textsc{#2}}}%
\fi}

\begin{document}

\title{Spectroscopic indication of suprathermal ions\\ in the solar corona}
\author{E. Lee$^{1}$, D. R. Williams$^{2}$, and G. Lapenta$^{1}$\\[5mm]
\small $^1$ Centrum voor Plasma-Astrofysica, KU Leuven \\
\small Celestijnenlaan 200B, B-3001 Leuven, Belgium \\[2mm]
\small $^2$ Mullard Space Science Laboratory, University College London \\
\small Holmbury St. Mary, Surrey, RH5 6NT, United Kingdom}
\date{15 February 2012}

\maketitle

\section*{Abstract}
{\small
{Solar spectroscopy traditionally assumes electrons and ions in coronal plasma exist in Maxwell--Boltzmann distributions, or thermal equilibrium.  Many studies in recent decades, however, have investigated the possibility of kappa distributions of coronal and solar wind particles far from Maxwellian and their resulting effects on collision, excitation, and ionization rates, heat transfer, waves, and instabilities.  Kappa distributions belong to a class of statistical equilibrium ensembles that are known solutions to the Boltzmann equation, in both collisional (Fokker-Planck) and collisionless (Vlasov) frameworks.  They have been observed \textit{in situ} in the solar wind and in planetary magnetospheres throughout the heliosphere.}
{Using spectroscopic data, we support the possibility of suprathermal distributions of coronal ions by fitting the equivalent kappa functions to their emission line profiles.}
{We fit different kappa and Gaussian model functions to line profiles of the strong \ion{Fe}{xv} line at 284.16 \AA, across two large-field spectroscopic rasters taken in a solar active region.  Both single- and double-component Gaussian models are applied, as well as two kappa models, one with a free width parameter allowing for and the other with a constrained width that precludes ``microturbulence.''  We then compare the goodness of fit of the computed best fits for each model.}
{The kappa distribution is a generalization, or superset, of the Maxwellian, so they are able to fit line profiles more precisely than a Gaussian.  In most of the data, the best-fit kappa model produces much lower residuals across the profile than any single Gaussian and sometimes double Gaussian.  Most importantly, the distribution of estimated kappa values is found to lie mostly in the low-$\kappa$ range, implying ion populations far from Maxwellian.  Even when the width is removed as a free parameter of fit, the kappa model is still able to fit the data credibly, again with low best-fit values of $\kappa$.}
{We find the shape of the \ion{Fe}{xv} line, in the vast majority of the data analyzed, to be indicative of a highly suprathermal ion population.}
}


\section{Introduction}

The solar atmosphere is composed of many different types of environments with wide ranges of magnetic field strength, density, temperature, abundances, scales of motion, and practically every other descriptor of plasma conditions.  While countless models have been proposed that describe plasma conditions in the various regions of the solar atmosphere from the photosphere to the corona, or from loop tops to coronal holes, no instrument to date has been able to make in-situ measurements to confirm the model predictions.  Nevertheless, much can be learned from remote sensing because many of the plasma conditions are encoded into the Sun's electromagnetic spectrum.


{\em Hinode}'s EUV Imaging Spectrometer (EIS) \citep{Kosugi07,Culhane07} provides valuable spectroscopic information, which can be deciphered to deduce the composition, density, and temperature of the plasma in the transition region and corona (0.04~MK $< T <$ 20~MK).  With spatial resolution as fine as 2~arcsec, spectral dispersion of 22.3~m\AA (with sensitivity to velocities down to 5~km~s$^{-1}$), it also provides the ability to differentiate plasma conditions across and off the solar disc and to detect waves and motions.



An important discovery due largely to observations by EIS was the identification of blueward Doppler shifts of EUV emission lines at open field lines close to active region boundaries and the likeliness of these regions to be sources of the slow solar wind \citep{Sakao07,Harra08}.  \citet{Baker09} identified the locations of quasi-separatrix layers (QSLs) in an active region and found them to correspond to blue shifts at the active region periphery.  Although the QSLs were computed at the photosphere, the study implied that the site of the magnetic reconnection driving the outflow was in the corona, since the blue shifts were strong for lines of hot ions in the corona (e.g. \ion{Fe}{xii}) but weak or red-shifted for those of cooler species \citep[cf.][]{DelZanna08}.

Blue shifts at active region boundaries have also been correlated with enhanced line broadening \citep{Doschek07, Doschek08, Hara08, Bryans10, BrooWarr11}.  Doppler line broadening associated with a Doppler shift of the same line was perhaps first reported by \citet{Nicolas82}; both were found to correlate positively with temperature in transition region lines.  The correlation between Doppler shift and broadening at active region boundaries is interesting because it links two processes that at first appear distinct and independent: the slow, steady upward flow out of an active region boundary and the so-called microturbulence that is conventionally assumed to cause excess line broadening.  Assuming turbulence exists, a difficult assumption to justify for a collisionless plasma, one possible explanation for the outflows is that magnetic reconnection is enhanced by the turbulence through anomalous resistivity \citep{Strauss88}.  

Atomic spectral lines are broadened by several effects, including radiation, opacity, the Stark effect, and Van der Waals damping \citep{Rutten}, but the dominant broadening effect in coronal lines are due to thermal motions of a hot, rarefied plasma.  A Gaussian profile can reflect the line-of-sight velocity distribution of a purely Maxwellian ion population, but more often than not, it is measured to be broader than what can be attributed to thermal motions, based on the peak ionization temperature.  Microturbulence is a convenient explanation for this excess, non-thermal broadening, but it is generally recognized as a ``fudge parameter'' \citep{Rutten}.



Whatever the mechanism, the same correlation between line shift and width was also noticed in regions of emerging flux (Harra et al. 2010).  In the case of flux emergence, the probability of magnetic reconnection is extremely high, as evidenced in numerous simulation studies \citep{FanGib04,Manchester04,Arch04,MarSyk08,MarSyk09,MacTagg09}.  However, while reconnection is known to cause both flows and particle acceleration, its role in creating actual turbulence in the corona is debatable.

Another explanation for line broadening is related to heterogeneity within the resolution element.  For example, within one or more pixels, a fraction of the plasma at active region boundaries could be propelled upwards in the form of small-scale jets, or spicules, against a background of stationary or slower-moving plasma.  Type-II spicules are believed to create secondary emission that can cause asymmetry of the profile, such as by a second Gaussian component enhancing the blue wing of the main Gaussian profile \citep{DePMcI09,McIDeP09,Peter10,TianMcIDeP11,MarSyk11}.  While type-II spicules are mainly found in the high chromosphere, they are believed to reach coronal heights and temperatures \citep{Judge12}.

\subsection{Kappa distributions} \label{sec:intro_kappa}

A revolutionary interpretation of non-thermal line broadening comes from \citet{Scudder92b}, which proposes that both the solar temperature inversion and excess line widths could be explained by ions and electrons out of thermal equilibrium, specifically, kappa-distributed populations.  Kappa distributions belong to a class of marginally stable equilibrium solutions of the Boltzmann equation, similar to that of the classical Maxwell--Boltzmann formalism, but whereas the latter describes the velocities of particles in thermodynamic equilibrium whose only interactions are local, elastic collisions, kappa distributions belong to a different class of statistical equilibria in which the constituent particles maintain long-range interactions, or non-extensivity.  Such non-local interactions, which distort what would otherwise be a thermodynamic equilibrium, can be in the form of broadband turbulent velocity correlations or by means of electromagnetic fields or waves \citep{RouDup80a,Hasegawa85}.  An immediate consequence is that the collision frequency becomes more complex than just a simple function of the average kinetic energy of the local plasma.  Therefore, particles in the high-energy tail of the distribution travel along nearly collisionless trajectories.  A comprehensive statistical mechanical theory, based on an alternative to Boltzmann's H-theorem---therefore maximizing a different, non-extensive entropy---has been developed by \citet{Tsallis88, Tsallis95}; the kappa distribution is an extension of the Tsallis entropy formalism \citep{PierLaz10}.

Most plasma kinetic theory is founded on Maxwell--Boltzmann equilibrium principles.  However, almost all plasma particle distributions measured in space are found to be suprathermal, having thicker tails than a Maxwell--Boltzmann distribution.  The kappa distribution is one such example, and it is often recognized in space plasma by its characteristic power-law tails \citep{Leubner02,Lapenta09}.  Many recent studies have reported evidence or theoretical predictions of kappa distributions in the solar wind \citep[see][and references within]{PierLaz10}, the exospheres of Earth and other planets \citep{PierLem96, Meyer01}, stellar coronae \citep{Scudder92a,Scudder92b,DorScud99,DorScud03}, and the interstellar medium \citep{GloeGei98,ElAwady10}.  While the kappa distribution is not the only distribution with power-law tails, it is a convenient description of space plasmas because it can be considered a generalization of the Maxwell--Boltzmann distribution.  In fact, in the limit of the parameter $\kappa \rightarrow \infty$, a Maxwellian is recovered.


Non-Maxwellian distributions in the solar wind were reported as early as 1968 by \citet{Montgomery68}.  They were also postulated to exist in the solar transition region by \citet{RouDup80a,RouDup80b}, which conjectured that electron and proton populations would necessarily form suprathermal populations with enhanced high-energy tails in regions of steep temperature gradient, such as the transition region, and that the high-energy tails could be maintained by electric fields.  Many subsequent studies have identified the kappa distribution as a likely model for solar atmospheric plasma, assessing their effect on ionization and recombination rates \citep{OwoScud83,Dzif92,Dzif98}, emission intensity \citep{AndRayVanB96,Dzif00,Dzif06}, excess line widths \citep{Scudder92b,Scudder94}, and coronal heating \citep{DorScud99,DorScud03}.

The three-dimensional kappa distribution function is given by

\begin{equation} \label{eqn:kappafunc}
f^\kappa(\bvec v) = A_\kappa \left( 1 + \frac{1}{\kappa} \frac{v^2}{v_0^2} \right)^{-\kappa-1}
\end{equation}
\noindent where
\[
A_\kappa = \left(\pi \kappa v_0^2 \right)^{-3/2} \frac{\Gamma(\kappa+1)}{\Gamma(\kappa-1/2)}
\]
\noindent with equivalent thermal velocity $v_0$, and $\kappa > \frac{1}{2}$ required for convergence.  Although the distribution is not technically thermal, an equivalent temperature $T$ can be defined, which can be related to the thermal velocity $v_0$, rms velocity $\bar{v}$, and mean energy $\langle \mathcal{E} \rangle$ by:
\[
v_0 = \sqrt{\frac{2\kappa-3}{\kappa} \ \frac{k_B T}{m}}
\]
\[
\langle \mathcal{E} \rangle = \frac{3}{2}k_B T = \frac{1}{2} m \bar{v}^2
\]

\noindent Although defining $v_0$ in physical terms requires $\kappa > \frac{3}{2}$, this stricter restriction on $\kappa$ is not required, except when relating to temperature.

Ion populations responsible for EUV emission in the corona are generally assumed to be Maxwellian, but in the next sections, we discuss the possibility of their being kappa-distributed, as evinced by their line profile shape.  In \S \ref{sec:data_analysis}, we describe the data and our models and methods for fitting the atomic line profiles; in \S \ref{sec:results} we detail the results of the different fits; and finally in \S \ref{sec:discussion} we discuss the implications of our results and the likelihood of kappa distributions of coronal ion populations.


\begin{figure*}[ht]
\centering
\includegraphics[width=0.45\textwidth]{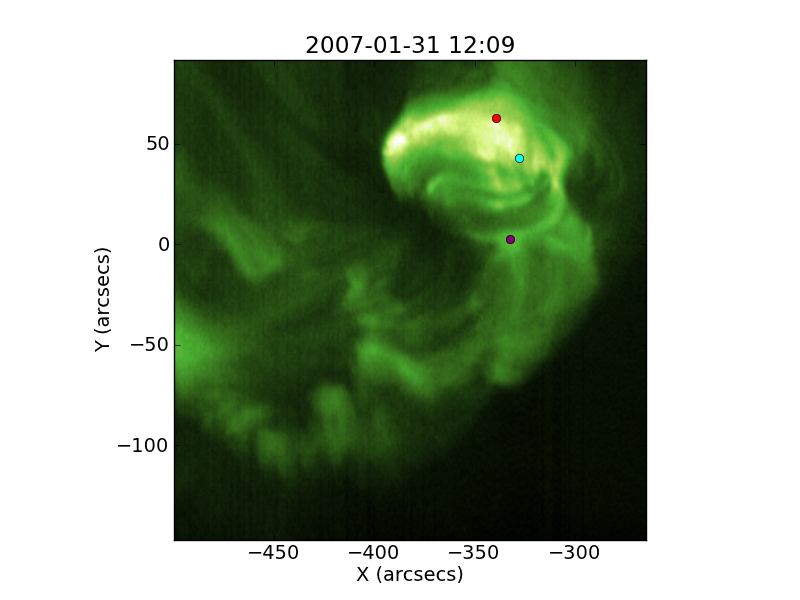}
\includegraphics[width=0.45\textwidth]{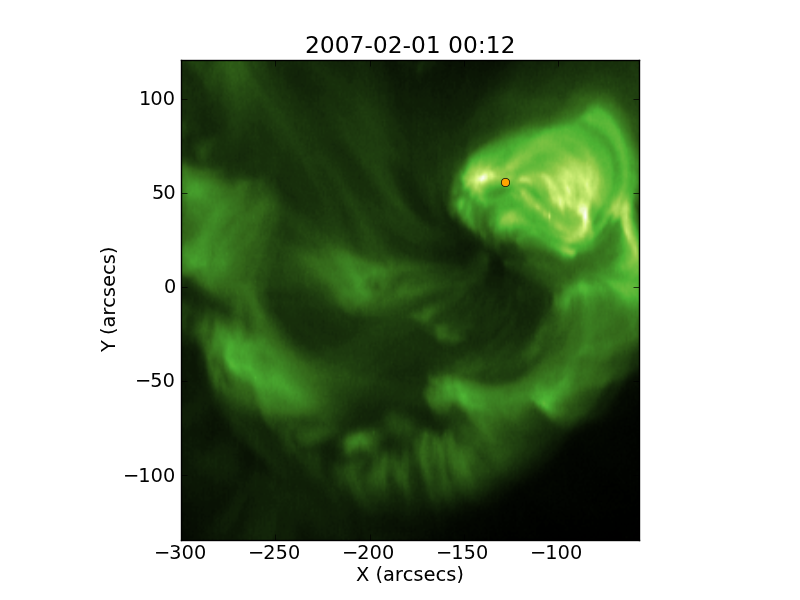}
\caption{\ion{Fe}{xv} 284.16~{\AA} line-integrated intensity for rasters at 31 January 2007 12:09:13 (left) and 1 February 2007 00:12:12 (right), with dots plotted where sample profiles are analyzed: A (cyan), B (red), C (purple), D (orange).}
\label{fig:FE_XV_intensity}
\end{figure*}

\begin{figure*}[ht]
\centering
\includegraphics[width=0.45\textwidth]{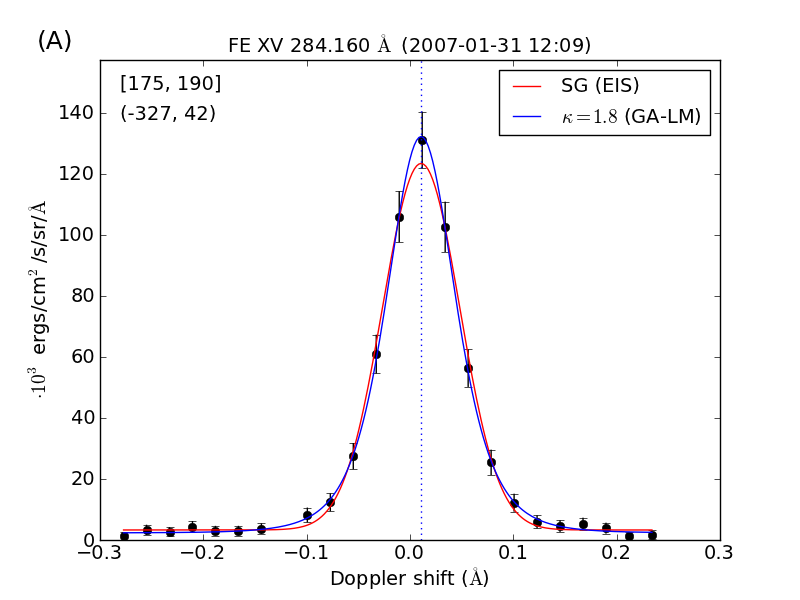}
\includegraphics[width=0.45\textwidth]{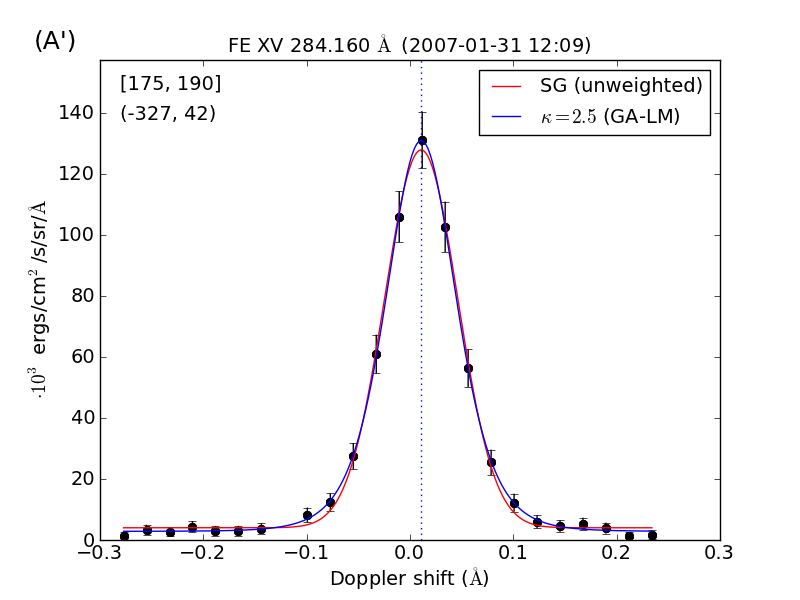}
\caption{Example of kappa (blue) and Gaussian (red) functions fit to \ion{Fe}{xv} emission line profiles at 284.16~{\AA} (2007-Jan-31 12:09:13).  Array coordinates are denoted in square brackets, and solar coordinates (arcsec) in curved brackets; see cyan dot in Fig.~\ref{fig:FE_XV_intensity}.  Left: best-fit curves using measurement error to studentize residuals ($\kappa=1.8$).  Right: best-fit curves for the same data, from unweighted fits ($\kappa=2.5$).}
\label{fig:kappa_SG_weighted_unweighted}
\end{figure*}


\begin{figure}[ht]
\centering
\includegraphics[width=0.45\textwidth]{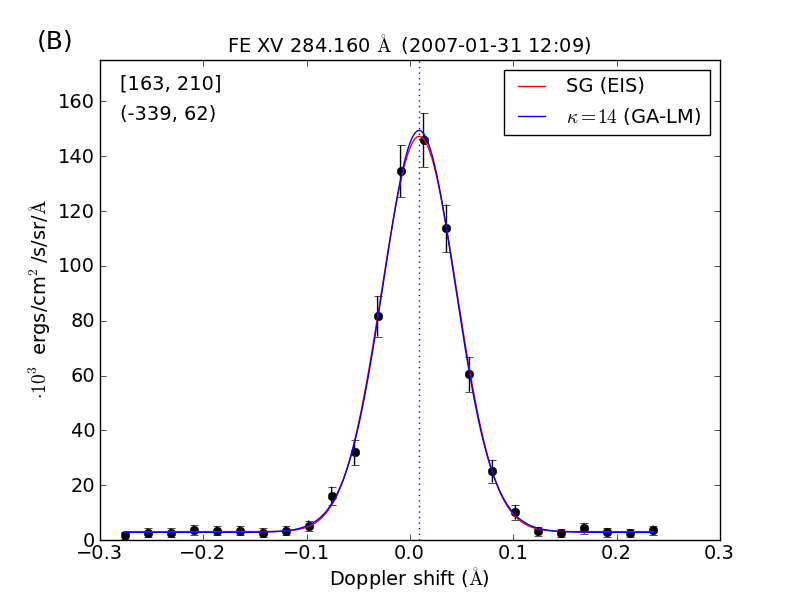}
\includegraphics[width=0.45\textwidth]{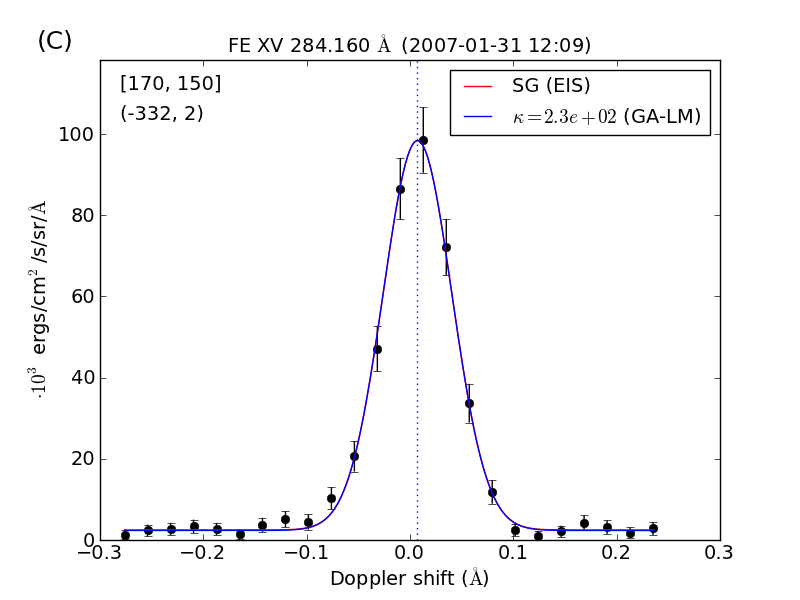}
\includegraphics[width=0.45\textwidth]{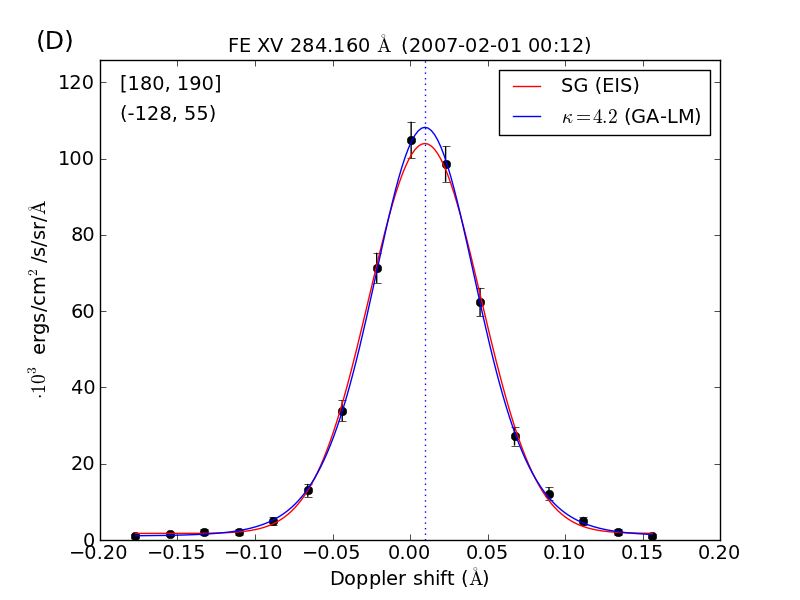}
\caption{Three more examples of kappa (blue) and single Gaussian (red) functions fit to \ion{Fe}{xv} emission line profiles at 284.16~{\AA}.  Profiles B and C, taken from the first raster, can be seen as red and purple dots, respectively, in Fig.~\ref{fig:FE_XV_intensity}, left; D is taken from the second and is marked in Fig.~\ref{fig:FE_XV_intensity}, right.  Best-fit curve parameters were obtained for all three profiles using measurement error to studentize residuals.  The computed values of $\kappa$ in B, C, and D are 14, 230, and 4.2, respectively.}
\label{fig:kappa_SG_BCD}
\end{figure}

\begin{figure*}[ht]
\centering
\includegraphics[width=0.45\textwidth]{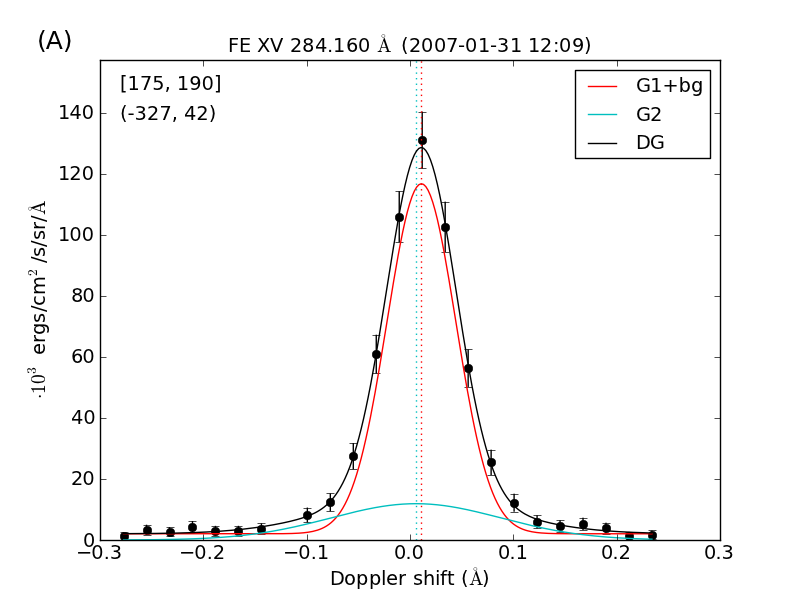}
\includegraphics[width=0.45\textwidth]{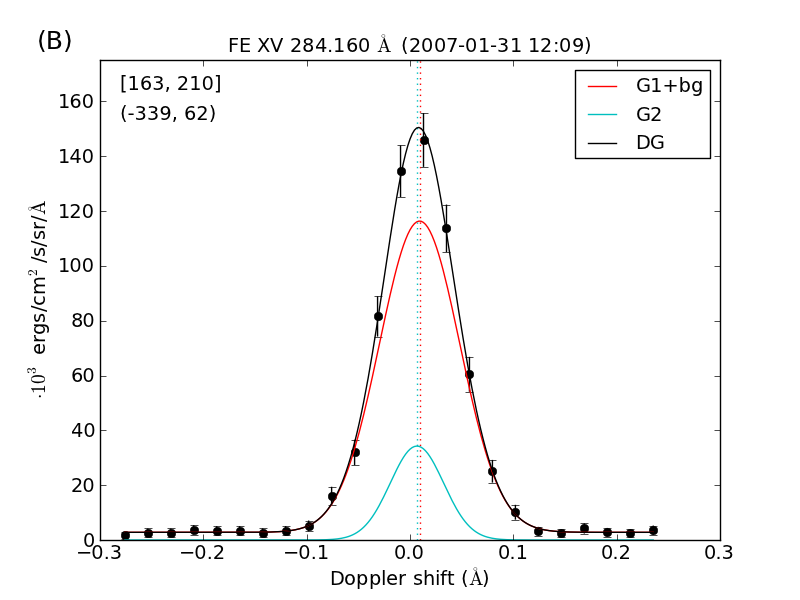}
\includegraphics[width=0.45\textwidth]{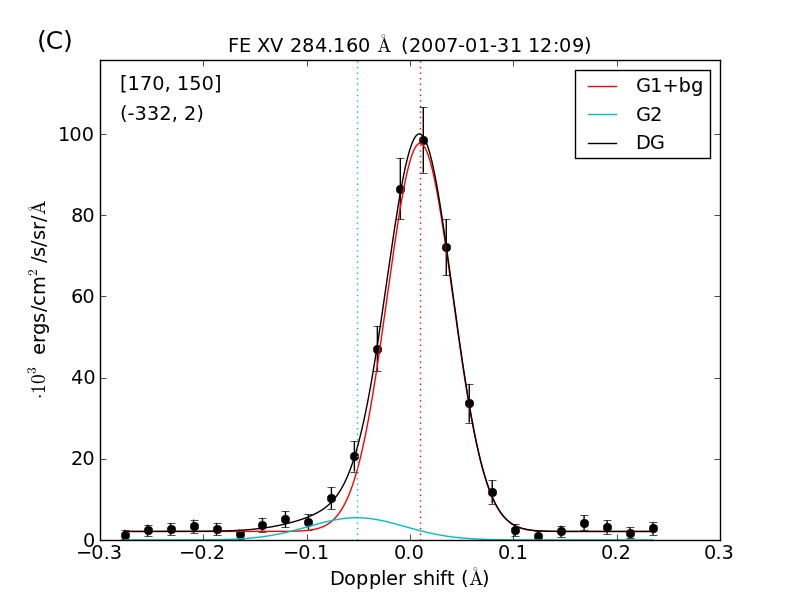}
\includegraphics[width=0.45\textwidth]{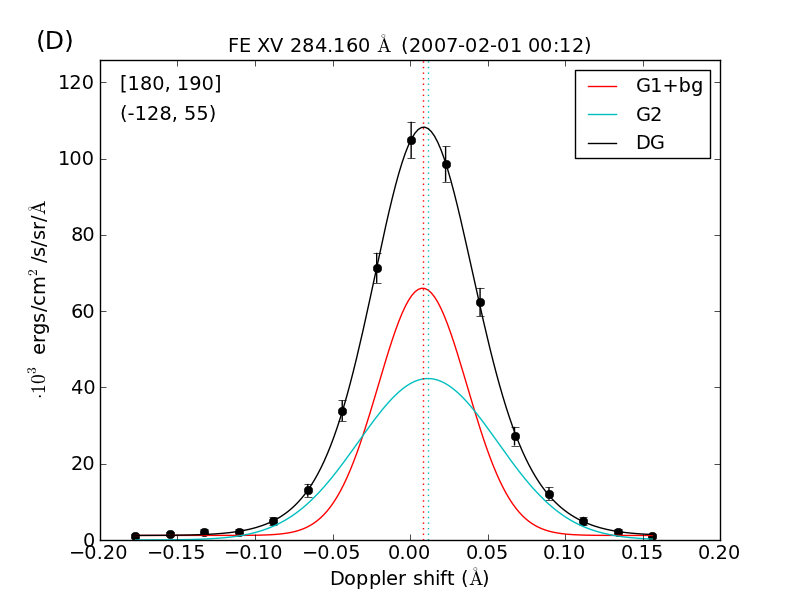}
\caption{The same examples of \ion{Fe}{xv} emission lines at 284.16~{\AA} (2007-Jan-31 12:09:13) as found in Fig.~\ref{fig:kappa_SG_weighted_unweighted} (left) and \ref{fig:kappa_SG_BCD}, instead fit to double Gaussian profiles, with both individual (cyan, magenta) and combined (black) components plotted separately.  The primary component (magenta) includes a flat continuum background.}
\label{fig:DG_ABCD}
\end{figure*}

\begin{table}[ht] \caption{Smoothness measure and reduced $\chi^2$ errors of (weighted) fits of the single Gaussian (SG), kappa ($\kappa$), double Gaussian (DG), and constrained-width kappa ($\kappa^*$) models applied to example profiles from 31 January 2007 12:09.}
\label{tbl:reduced_chisq}
\centering
\begin{tabular}{c c c c c c}
  \hline
  Profile & Rs & $\chi_r^2$(SG) & $\chi_r^2$($\kappa$) & $\chi_r^2$(DG) & $\chi_r^2$($\kappa^*$) \\
  \hline\hline
  A & 1.61 & 0.721 & 0.293 & 0.290 & 0.298 \\
  B & 1.72 & 0.271 & 0.239 & 0.257 & 0.505 \\
  C & 1.49 & 0.638 & 0.633 & 0.414 & 0.776 \\
  D & 1.59 & 0.909 & 0.274 & 0.264 & 0.424 \\
  \hline
\end{tabular}
\end{table}

\begin{table}[ht] \caption{Smoothness measure and averaged unbiased residuals for (weighted) fits of the single Gaussian (SG), kappa ($\kappa$), double Gaussian (DG), and constrained-width kappa ($\kappa^*$) models applied to example profiles from 31 January 2007 12:09.}
\label{tbl:unbiased_RSS}
\centering
\begin{tabular}{c c c c c c}
  \hline
  Profile & Rs & $\bar{\epsilon}$(SG) & $\bar{\epsilon}$($\kappa$) & $\bar{\epsilon}$(DG) & $\bar{\epsilon}$($\kappa^*$) \\
  \hline\hline
  A & 1.61 & 0.502 & 0.204 & 0.210 & 0.185 \\
  B & 1.72 & 314 & 0.287 & 0.288 & 0.625 \\
  C & 1.49 & 0.318 & 0.323 & 0.194 & 0.429 \\
  D & 1.59 & 0.395 & 0.158 & 0.124 & 0.284 \\
  \hline
\end{tabular}
\end{table}


\section{Data Analysis} \label{sec:data_analysis}

\subsection{Data}
The {\em Hinode}/EIS observations that we analyze are of active region 10940, a simple $\beta$ region which crossed the Sun in early 2007 (see Fig.~\ref{fig:FE_XV_intensity}).  These and other EIS data from this region have previously been analysed by \citet{Warren:2007p407,Warren:2008p2251,Ko:2009p5820,Wang:2009p8915}.  Region 10940 produced only a handful of small flares during its passage across the solar disc, the last and largest (class C3.5) occurring on 2009 Jan 29.  The raster data we investigate here began at 2007 Jan 31 12:09 and 2007 Feb 1 00:12~UT, and were built up by scanning the 1'' slit over an area of approximately 250''~$\times$~250'' that covered most of the active region.  The data contain a number of coronal emission lines, including, but not limited to, lines from \ion{Fe}{viii} to \ion{Fe}{xvi}.  (The original data are available via http://bit.ly/EISkappaData .)

These observations were calibrated in IDL using the standard routine {\tt EIS\_PREP} available through SolarSoft \citep{Freeland:1998p493}, removing the effects of dark current, CCD read-out bias, and spikes of anomalous charge in individual CCD pixels due to warm and hot pixels and cosmic rays.  We also calibrate the data to specific radiative intensity units and implement the absolute wavelength scale calibration by \citet{Kamio:2010p9683} to minimize the effects of thermal distortion of the instrument around {\em Hinode}'s orbit, as well as removing the effects of slit tilt (with respect to the cross-dispersion direction) and the wavelength-dependent offset of the field-of-view in the direction of the slit.

With the data calibrated and corrected for instrumental effects, we then applied a filter to eliminate ``jagged'' lines from the data, which would be ill fit by a unimodal (single-peaked) model function, by evaluating a smoothness measure, $R_s$, which is a ratio of Sobolev seminorms, defined in Appendix~\ref{app:smoothness}.  $R_s$ decreases with the jaggedness of the profile, which is to say that it is higher for smoother, less oscillatory lines.  We chose an arbitrary cut-off value for $R_s$, fitting only those for which $R_s>1$.

Among the spectral lines available in the data, we chose to analyze the \ion{Fe}{xv} (2~MK) line at 284.16~{\AA}, which is considered a clean line, in general \citep{Young:2007p3278}, and also by our smoothness criterion $R_s$.  Even still, a large proportion of the profiles in the first raster had $R_s<1$, due to low intensity and signal-to-noise ratio, while in the second, most of the profiles were smooth enough for unimodal fitting; this is likely due to the longer exposure time of the second raster.

\subsection{Profile fitting} \label{sec:profile_fitting}

%



We applied the standard Gaussian profile fitting routines provided in SolarSoft ({\tt EIS\_AUTO\_FIT}) to compute for each pixel the amplitude, centroid, and width of the emission line, as well as the background continuum level.  The weighted least-squares fit is implemented as a local iterative algorithm, specifically the Levenberg--Marquardt (LM) nonlinear least-squares routine provided in IDL through MINPACK \citep{Markwardt09}.  We assumed a flat (spectrally uniform) background continuum and a single Gaussian profile, given the cleanliness of the line.  The model function can be written

\begin{equation} \label{eqn:gaussfunc}
I(\lambda; b, A, \mu, \sigma) = b + A \exp \left[- \frac{(\lambda-\mu)^2}{2 \sigma^2} \right]
\end{equation}

\noindent with background $b$, amplitude $A$, centroid $\mu$, and width $\sigma$.



In addition to a Gaussian model, corresponding to a Maxwellian plasma with normally distributed non-thermal motions, we also considered a kappa-distribution model, with function:
\begin{equation}
I(\lambda; b, A, \mu, \sigma, \kappa) = b + A \left[ 1 + \frac{(\lambda - \mu)^2}{\kappa \sigma^2} \right]^{-\kappa-1}
\end{equation}
\noindent where $\kappa$ is an additional fit parameter controlling the shape of the profile but also affecting its width.

For the kappa model, each profile was fit by a two-step process:  first a genetic algorithm (GA) method for global optimization called PIKAIA \citep{Char95,CharKnapp95} was used to produce an estimate for the parameters, and then a second, local optimization was performed, using the GA output as an initial guess.  The combination of the GA method and local refinement by LM is commonly used for spectral line fitting, especially for complex profiles, such as double Gaussians \citep[e.g.,][]{Peter10}.


The generality of the kappa function is better able to capture excess emission in the tails, compared to a single Gaussian.  However, a second, wider Gaussian component could also account for excess wing emission.  For the purpose of comparison, we also performed double Gaussian fits on the same data, using also the combination of GA global optimization and local refinement by LM.  The double Gaussian function is given by:


\begin{eqnarray} \label{eqn:DGfunc}
I(\lambda; b, A_1, \mu_1, \sigma_1, A_2, \mu_2, \sigma_2) = & b + A_1 \exp \left[- \frac{(\lambda-\mu_1)^2}{2 \sigma_1^2} \right] \nonumber\\
 & + A_2 \exp \left[- \frac{(\lambda-\mu_2)^2}{2 \sigma_2^2} \right]
\end{eqnarray}


It is sometimes difficult to distinguish between a kappa function and a combination of Gaussians, especially when both the peak and the tails are represented by just a few points.  With better precision in tail measurements, however, they can be clearly distinguished.  In the case of fast solar wind electrons, for example, measurements of velocity distributions by \textit{Ulysses} show indubitably that the tails cannot be fit by multiple Gaussians but are fit very convincingly by a kappa function \citep{MakPierRil97}.  Nevertheless, for a sparsely sampled distribution or curve, the difference between the two models is minute, and in addition, the double Gaussian model is generally likely to minimize the residual better because it contains two more degrees of freedom than the kappa model and allows for asymmetry of the wings.



The least-squares method used for fitting each of the model functions $I(\lambda)$ minimizes the residual sum of squares (RSS), where the residuals are the vertical offsets between the data and the best-fit curve.  Except where noted (specifically, for Profile A$^\prime$ only), the residuals used in the fit were ``studentized,'' or weighted by estimated measurement error.  For each profile fit, we report the reduced chi-square error, which is the studentized RSS divided by the number of degrees of freedom:
\begin{equation} \label{eqn:reduced_chisq}
\chi_r^2 \left(I(\lambda), y, w\right) = \frac{1}{N-n_p} \ \sum_{i=1}^{N} \frac{\left(I(\lambda_i)-y_i\right)^2}{w_i^2} 
\end{equation}
\noindent where $N$ is the number of sampling points across the $\lambda$ profile, $n_p$ is the number of parameters in the model, and $y$ and $w$ are the intensity measurements and sampling errors, respectively.  Equation \ref{eqn:reduced_chisq} is in fact the quantity being minimized by the weighted fitting algorithms.

In addition to $\chi_r^2$, we also report the averaged unweighted residuals to characterize the \textit{unbiased} goodness of fit:
\begin{equation} \label{eqn:averaged_residual}
\bar{\epsilon} \left(I(\lambda), y, w\right) = \frac{1}{N} \ \sum_{i=1}^{N} \left(I(\lambda_i)-y_i\right)^2
\end{equation}

\noindent It is important to emphasize to the reader that although $\bar{\epsilon}$ is the quantity which is minimized in an \textit{unweighted} fit, it is computed for all \textit{weighted} fits, in addition to $\chi_r^2$.


\begin{figure*}[ht]
\centering
\includegraphics[width=0.45\textwidth]{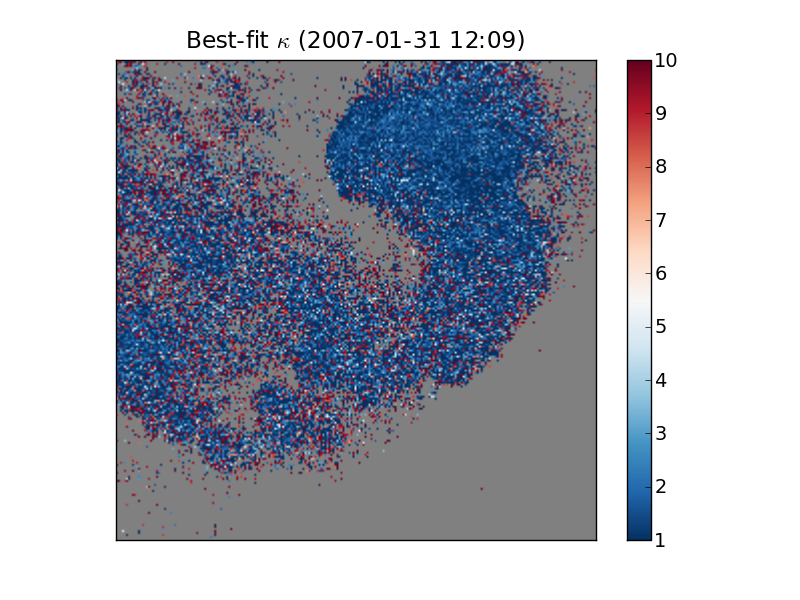}
\includegraphics[width=0.45\textwidth]{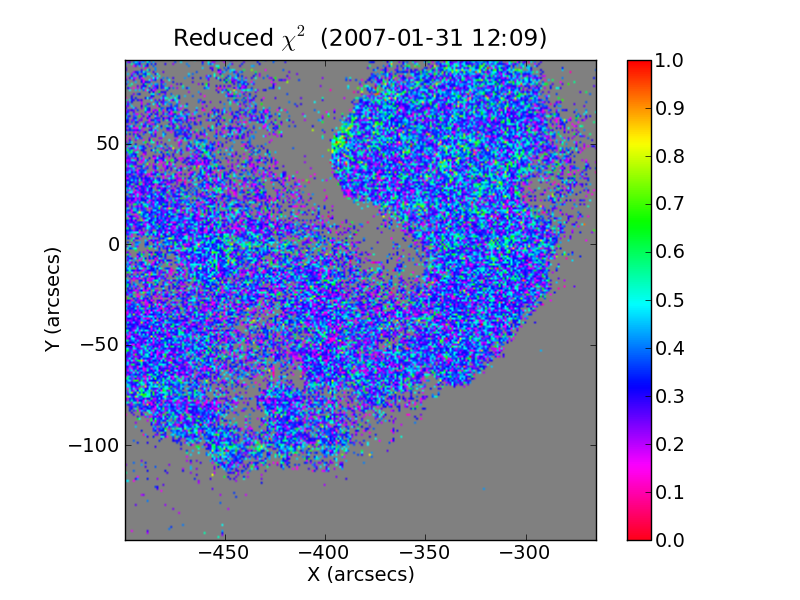}
\includegraphics[width=0.45\textwidth]{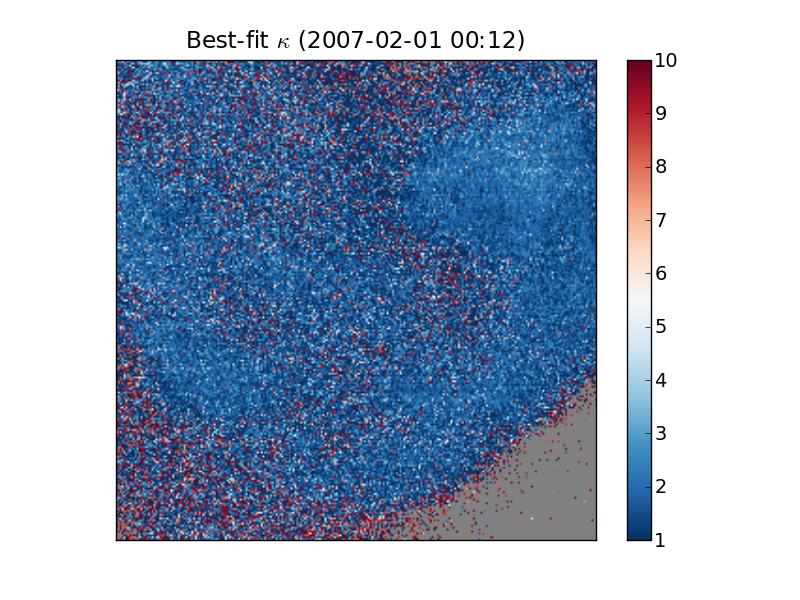}
\includegraphics[width=0.45\textwidth]{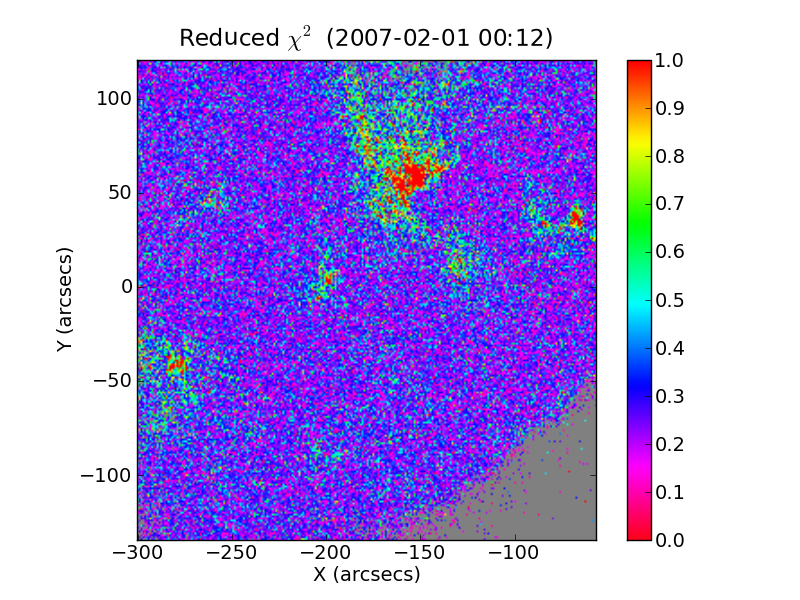}
\caption{Kappa maps (left) and reduced $\chi^2$ errors (right) for the two rasters.  Gray-masked areas contain profiles deemed too jagged for unimodal fitting, i.e., $R_s<1$.}
\label{fig:kappa_maps}
\end{figure*}

\begin{figure*}[ht]
\centering
\includegraphics[width=0.45\textwidth]{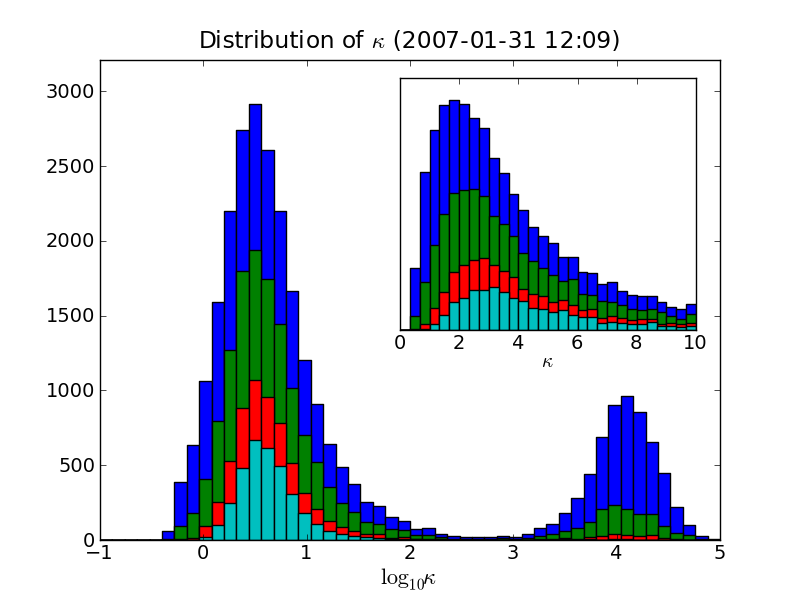}
\includegraphics[width=0.45\textwidth]{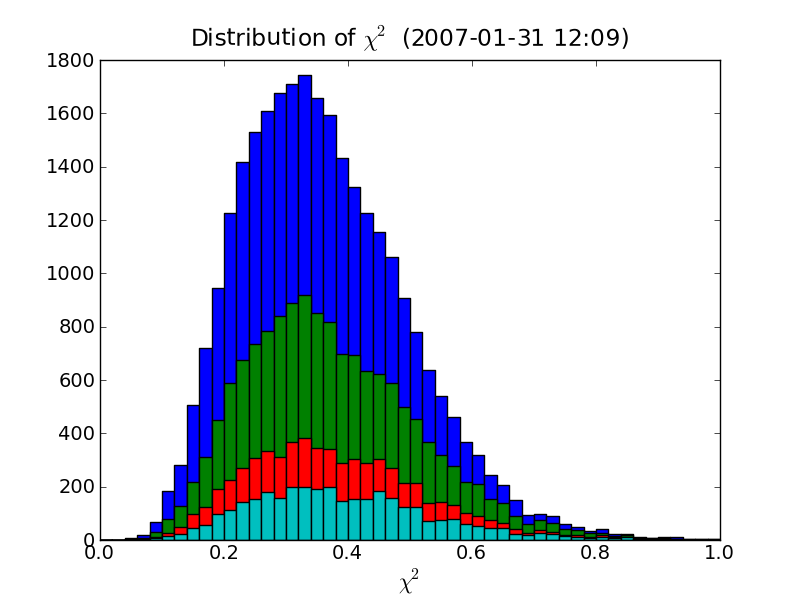}
\includegraphics[width=0.45\textwidth]{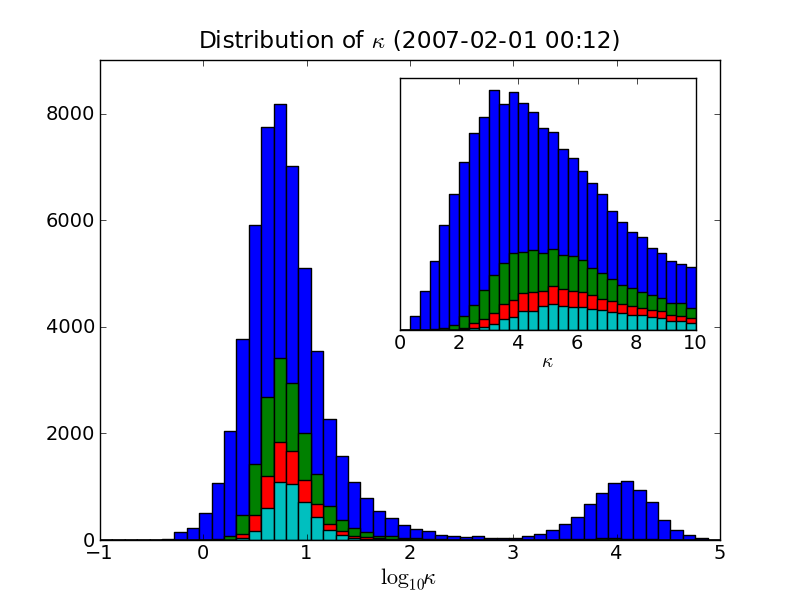}
\includegraphics[width=0.45\textwidth]{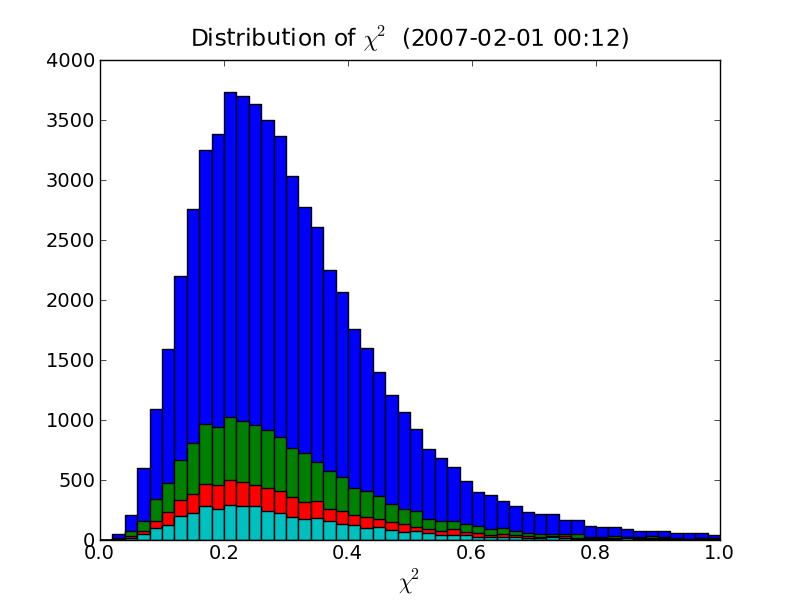}
\caption{Histograms of $\log \kappa$ (left), with linear-scale histograms of $\kappa$ inset, and histograms of $\chi^2$ (right).  For each group, the largest histogram represents all data, and each smaller histogram represents profiles whose intensity is above 20\%, 30\%, and 40\% of maximum intensity, in order of decreasing area.}
\label{fig:kappa_histograms}
\end{figure*}

\begin{figure*}[ht]
\centering
\includegraphics[width=0.45\textwidth]{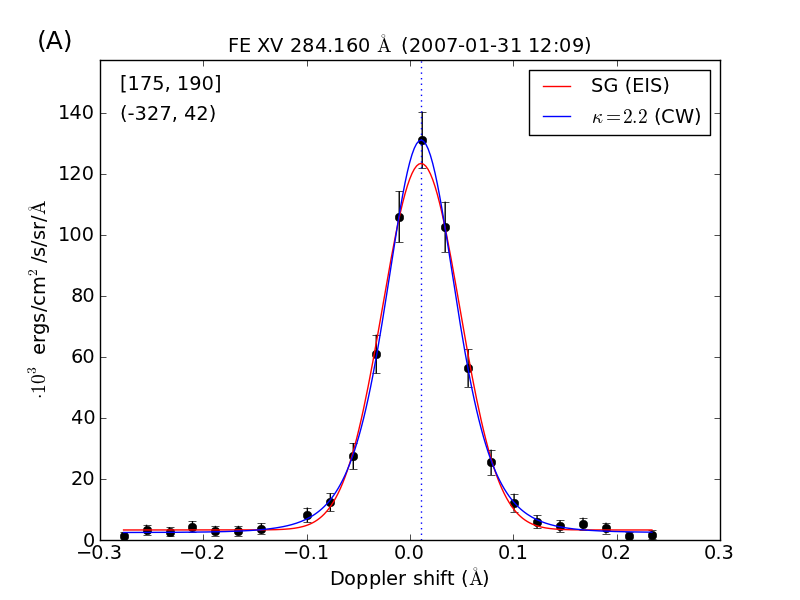}
\includegraphics[width=0.45\textwidth]{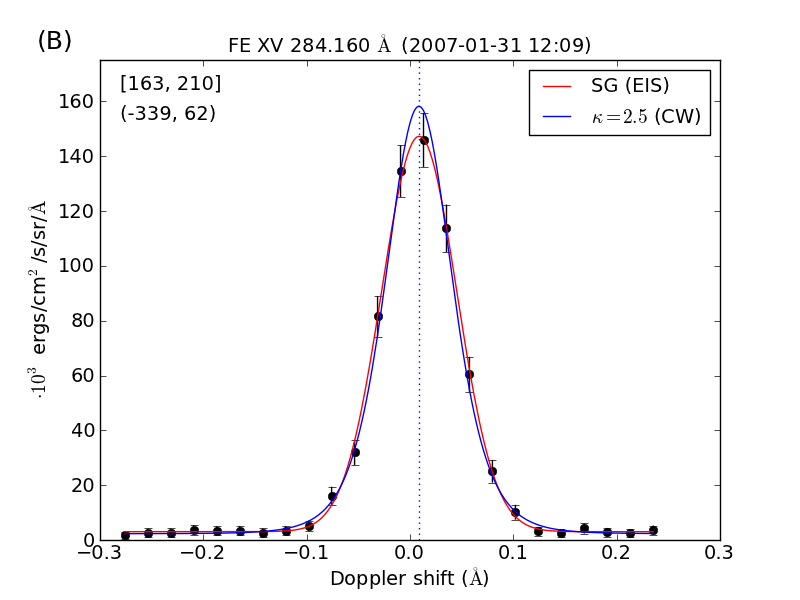}
\includegraphics[width=0.45\textwidth]{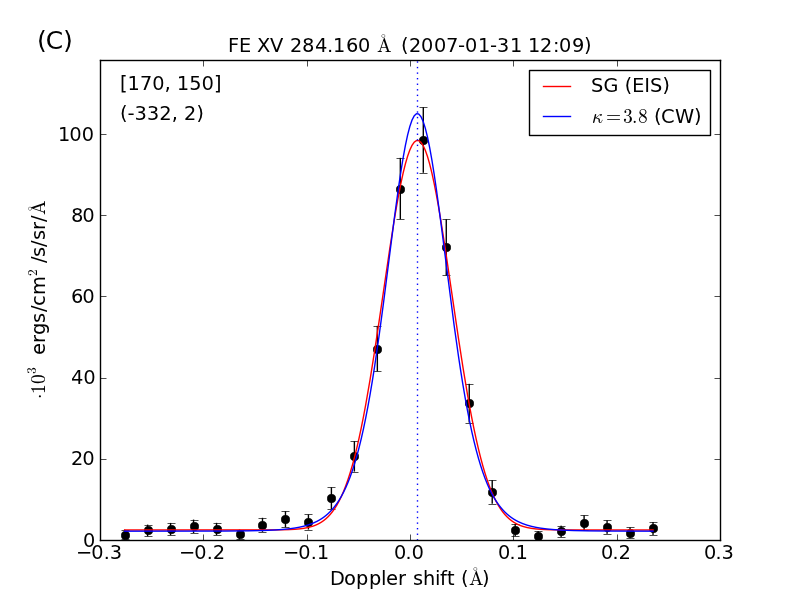}
\includegraphics[width=0.45\textwidth]{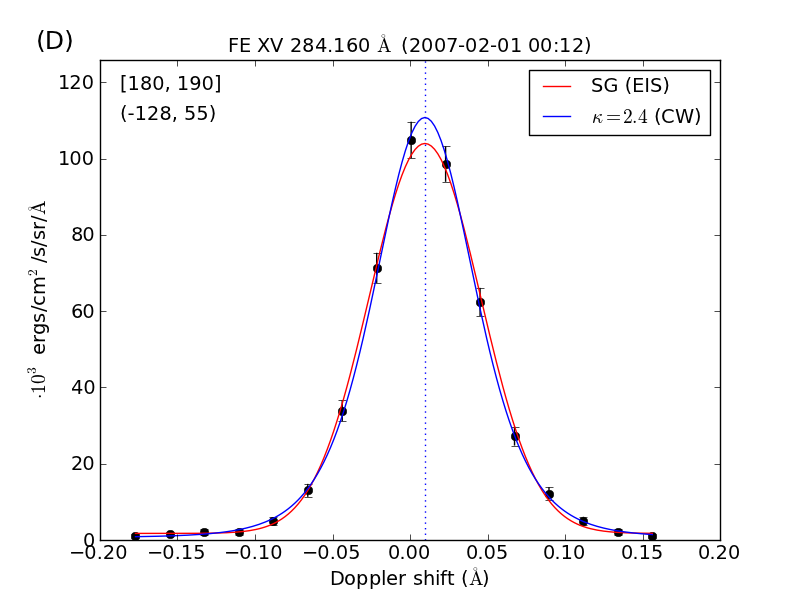}
\caption{The same examples of \ion{Fe}{xv} emission lines at 284.16~\AA \ (2007-Jan-31 12:09:13) as found in Fig.~\ref{fig:kappa_SG_weighted_unweighted} (left) and \ref{fig:kappa_SG_BCD}, instead fit to constrained-width kappa profiles.  Profiles A, B, C, and D have best-fit $\kappa$ values of 2.2, 2.5, 3.8, and 2.4, respectively.}
\label{fig:kappa_cw_ABCD}
\end{figure*}

\begin{figure*}[ht]
\centering
\includegraphics[width=0.45\textwidth]{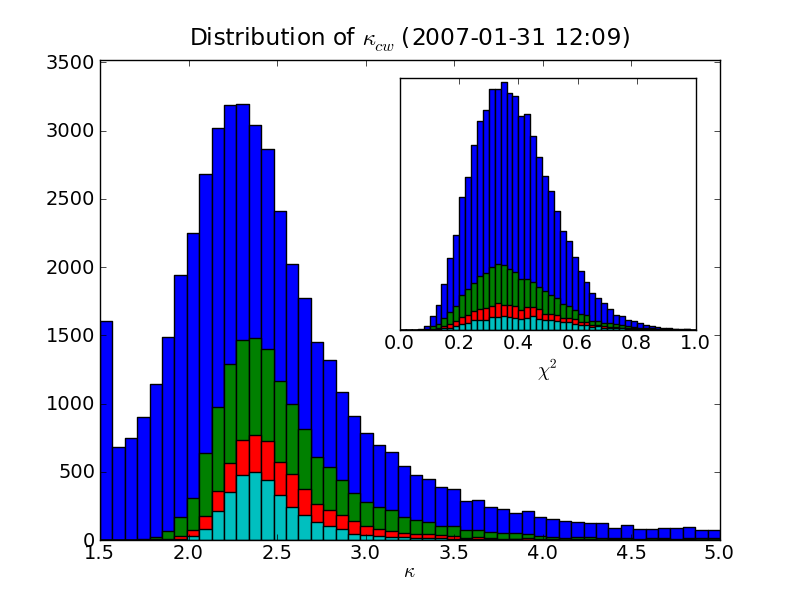}
\includegraphics[width=0.45\textwidth]{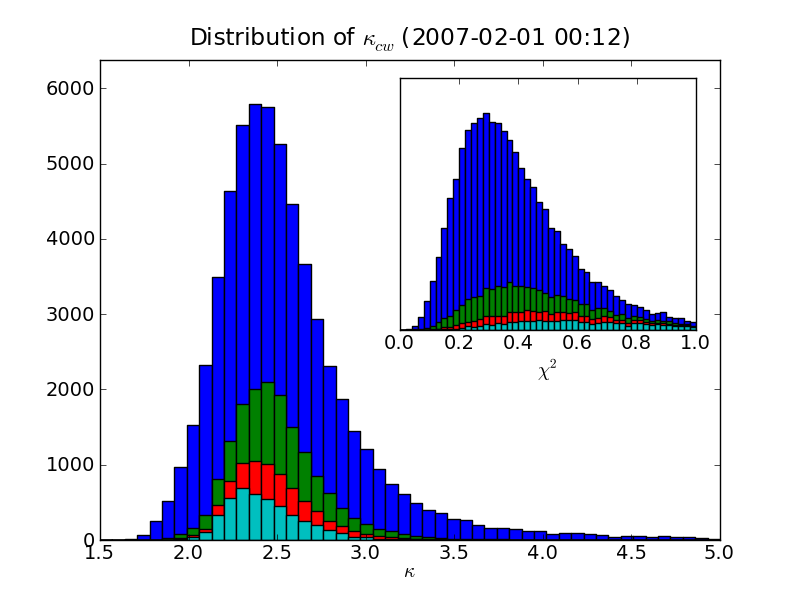}
\caption{Histograms of kappa (main) and $\chi^2$ (inset) for constrained-width fitting. Note that the horizontal ($\kappa$) scale is linear here, since no second hump ($\kappa>10^3$) appears in the distributions.}
\label{fig:kappa_cw_histograms}
\end{figure*}

\section{Results} \label{sec:results}

For most profiles in the data, a single Gaussian (hereafter, SG) fit shows a clear tendency to miss both the peak and the tails of the emission profile.  In fact, for the unweighted fit, the best fit reflects a competition between the peak residual and the tail residuals (Fig.~\ref{fig:kappa_SG_weighted_unweighted}, right).
When the residuals are weighted, however, the algorithm chooses a different best-fit curve, putting more importance on points for which the measurement error is smaller.  The left panel of the same figure illustrates that the weighted least-squares method allows for a greater discrepancy at the peak than in the wings, compared to the unweighted fit.  To reiterate, with the exception of the fit of Profile A$^\prime$ in Fig.~\ref{fig:kappa_SG_weighted_unweighted}, right, all parameters were found by \textit{weighted} least-squares, \textit{viz.}, minimizing $\chi^2$.

Kappa profiles can differ significantly from SG profiles, but they can sometimes be indistinguishable from Gaussians even at finite, though high, values of $\kappa$. 
The best-fit kappa curves are also different between the two panels in Fig.~\ref{fig:kappa_SG_weighted_unweighted}; the two methods in fact admit different best-fit values of $\kappa$.  Nevertheless, in both cases the kappa curves are better able than their SG counterparts to capture both the peak and the tails of the profile.

While most of the data tend to be fit best by low values of kappa, we indeed see some data points for which the kappa and SG curves are similar or even indistinguishable.  Profile B (Fig.~\ref{fig:kappa_SG_BCD}, top) illustrates that the $\kappa \approx 14$ and the SG fits are slightly different but that both fits are credible and well within the 1$\sigma$ error bars of the data.  In Profile C, the SG is completely overlain by the kappa function, and the two are distinguished only by their residuals.  Profile D was chosen from the core of the bright active region in the second raster (Fig.~\ref{fig:FE_XV_intensity}).  The only marked difference between D and the other three profiles is that the spectral window is shorter, and therefore there are fewer data points, although the spectral resolution is the same.  The fewer data points, however, translates into fewer degrees of freedom and thus slightly higher reduced $\chi^2$ values.

Table \ref{tbl:reduced_chisq} lists for each model the reduced $\chi^2$, a measure of the goodness of fit.  The fact that almost all the reduced $\chi^2$ values are smaller than 1 implies that all models overfit the data.  The $\chi^2$ errors for the kappa model, $\chi_r^2(\kappa)$, are consistently lower than those of the SG fit, and their difference is greatest, naturally, when the $\kappa$ value is low.

While the weighted least-squares algorithm is designed to minimize $\chi^2$, another measure of the goodness of fit is the average unbiased residual $\bar{\epsilon}$, (Table~\ref{tbl:unbiased_RSS}).  This quantity can be considered an absolute measure of error, in contrast to $\chi^2$, which is a more relative measure.  Comparing $\bar{\epsilon}$(SG) and $\bar{\epsilon}(\kappa)$, we see that the (weighted) kappa fits again produce smaller absolute residuals than the SG, with the exception of Profile C, where $\bar{\epsilon}(\kappa)$ is slightly higher.  We discuss possible reasons for this discrepancy in \S \ref{sec:discussion}.

We can also compare these fits to those attained by the double Gaussian model (Fig.~\ref{fig:DG_ABCD}).  The double Gaussian (hereafter, DG) best-fit curve, in general, produces smaller unbiased RSS, though not always.  With three extra parameters compared to the SG model, or two more than the kappa model, it is hardly surprising that the DG should be better equipped to fit the data.  The most significant advantage that the DG has over the SG or kappa models is that it is the only model that allows for asymmetry.  This important difference has been the subject of many papers, \citep[e.g.,][]{Peter10,MarSyk11}.

The DG model, however, has some general disadvantages:  the secondary component can have arbitrary width, which can fluctuate wildly over neighboring pixels, and the wider component produces a different curvature in the tails of the velocity distribution than might actually exist \citep[cf.][]{MakPierRil97}.  In our data set, we can see that, for some pixels, while the second Gaussian component can improve upon the fit of the SG, a DG does not necessarily minimize the residuals better than the kappa fit.  Comparing the fits of Profile A (Figs.~\ref{fig:kappa_SG_weighted_unweighted} and \ref{fig:DG_ABCD}), we note that only the kappa model is able to capture the peak, while both SG and DG slightly miss it.  Nevertheless, the difference between the kappa and DG fits is arguably insignificant, especially given that both can already be considered overfit by the reduced $\chi^2$ measure.  In the case of an asymmetric profile, such as Profile C (Fig.~\ref{fig:DG_ABCD}, lower left panel), the DG fit can be significantly better than the SG and kappa, which are constrained to be symmetric.

Another peculiarity of the DG fit is that the amplitude of the second component can be as arbitrary as its width; it is only a matter of algorithmic design to determine which is the primary component and which is secondary.  Profile D is an example of a DG fit in which the two components are of approximately equal importance (they have approximately equal area underneath).  The coexistence of two components with similar intensity but different widths is certainly possible but requires a model for understanding.

\subsection{Distribution of kappa.}

We fit profiles for each of the $240 \times 240$ pixels in the first raster and $256 \times 256$ in the second, excluding only those profiles whose smoothness measure was below our cutoff value ($R_s<1$) or for which the fitting algorithm failed to converge.  The spatial distribution of best-fit values of kappa can be seen in Fig.~\ref{fig:kappa_maps}, alongside a map of $\chi^2$.  The overwhelming tendency of the best fits is towards $\kappa$ values in the range $1<\kappa<5$, indicating markedly non-Maxwellian populations of ions almost everywhere.  There is some scatter of $\kappa$ values between neighbouring pixels, but it tends to be reduced in regions of high intensity, such as the core of the active region, in the upper portion of each image (cf. Fig.~\ref{fig:FE_XV_intensity}), where values of $\kappa < 5$ persist in the first raster and $\kappa < 6$ in the second.  The $\chi^2$ values tend to be low almost everywhere, except at the eastern edge of the bright region, where the intensity is low and upflow is strong (Doppler velocity image not shown).

Fig.~\ref{fig:kappa_histograms} presents histograms of best-fit values of $\kappa$ for the two rasters.  Each main panel shows histograms of $\log \kappa$ that are colored according to different levels of thresholding by intensity.  The largest (blue) histograms represent all the pixels for which a kappa fit was performed.  The three smaller histograms (green, red, cyan) represent pixels whose intensity is above 20\%, 30\%, and 40\% of the maximum intensity (in order of decreasing area).  The insets show similar sets of histograms but for a linear $\kappa$ scale.

From the logarithmic histograms, we can see that the computed values span a wide range.  Without filtering by intensity, we see two definite humps in the distribution.  However, as we filter by intensity, we see that the brightest pixels do not tend to lie on the high-$\kappa$ end of the distribution.  In fact, the second raster, which is very clearly superior in data quality due to its longer exposure time, has almost no pixels in the high-kappa domain, above the 20\% of maximum intensity level.  The absence of high-$\kappa$ values from high-intensity data could be related to its higher signal-to-noise ratio.  Alternatively, very high incident flux could cause detector saturation and therefore distort the shape of the profile.  However, no such saturation can be found in our data, and furthermore saturation would tend to flatten the profiles, whereas low values of $\kappa$ are generally associated with leptokurtosis, or steeper peaks and thicker tails.  Therefore, this would be a weak explanation, especially considering that the distributions of $\chi^2$ are not shifted to higher values by the filtering.  We must also consider the possibility of unresolved blends with \ion{Fe}{xv} at 284.16~{\AA}.  This line is known to be a clean line in the solar corona, but in the quiet Sun, there is a line due to \ion{Al}{ix} at 284.03~{\AA} \citep{Young:2007p3278}.  Although a blend between these two lines is possible along the line of sight, the \ion{Al}{ix} line is weak ($<$10\% of the \ion{Fe}{xv} line intensity; cf.~\citealt{Brown:2008p2270}).  It also formed at a lower temperature and occurs at a substantially shorter wavelength than that of \ion{Fe}{xv}, so that such blending would likely appear asymmetric.

The advantage of a logarithmic histogram is that the entire range, including the second hump, is visible.  However, it is somewhat misleading because the bins are not equally spaced but rather logarithmically separated.  Therefore, we also provide linear histograms of $\kappa$ between 0 and 10, with 0.5 being the minimum value of $\kappa$.  In both panel insets of Fig.~\ref{fig:kappa_histograms}, left, but especially that for the second raster (bottom left), we see that the distribution shifts to higher values of $\kappa$ with progressive intensity filtering.  In general, higher-intensity profiles tend to have fewer $\kappa$ values in both the low and high extremes of the scale, but that the tendency is for profiles to have low rather than high values of $\kappa$ when this filtering is applied.

\begin{figure}[ht]
\centering
\includegraphics[width=0.45\textwidth]{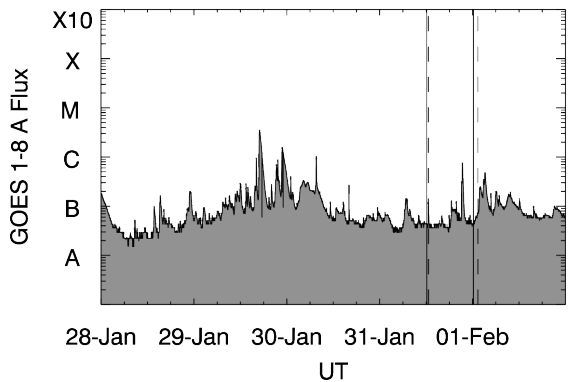}
\caption{GOES Sun-integrated soft X-ray flux starting when AR~10940 became visible on the solar disk. Solid/dashed lines indicate the start/end times of the EIS raster data analyzed.}
\label{fig:goes}
\end{figure}

A clear difference can be seen in the distribution of $\kappa$ between the two rasters: each of the intensity-filtered histograms for the first raster peaks at lower values of $\kappa$ than for the second.  One possible explanation is that most of the new flux in the active region has already emerged by the time of the first raster, while the histograms for the second raster could reflect a relaxation of the plasma.  Fig.~\ref{fig:goes} shows how activity has diminished by the start time of the first raster, while the next flare produced by the region comes after the end of the second (the intervening B-class flare does not come from this region).  Magnetic reconnection, a possible candidate for the energization of the ions in the tails, would be faster and more energetic during flux emergence, as the new and existing flux systems adapt topologically to rapid change in magnetic field strength and geometry.  Relaxation to high values of $\kappa$ could suggest diminishing levels of reconnection and/or reconnected flux between the first and second rasters.  Alternatively, it could just be that the first raster had noisier data; this possibility would be supported by the fact that more pixels had $R_s < 1$ and were thus filtered out in the first raster (Fig.~\ref{fig:kappa_maps}).  However, the relative shift in $\chi^2$ between the two distributions (Fig.~\ref{fig:kappa_histograms}, right panels) is \textit{not} indicative of poorer fits in the first raster but is rather the consequence of having more degrees of freedom in the denominator.

\subsection{Constrained width.}

\citet{Scudder92b} explains that the non-thermal excess widths of spectral lines, with respect to thermal broadening, could be explained by the difference in shape of kappa distributions compared to Maxwellians, without resorting to \textit{ad hoc} notions of turbulence.  The width of a kappa profile due to ``thermal'' broadening (collisional spread) is:
\[
\Delta \lambda = \frac{\lambda v_0}{c} = \frac{\lambda}{c} \sqrt{\frac{2\kappa-3}{2\kappa} \ \frac{2 k_B T}{m}}
\] 
\noindent The lines are also broadened by natural and instrumental effects.  We compute instrumental broadening from the calibration routines provided in SolarSoft but neglect natural broadening, as it is expected to be small.  We then fit the profiles, as described in \S \ref{sec:profile_fitting}, but with one fewer parameter:  $\sigma$ is now bound to $\kappa$ and dependent on both the peak temperature of formation ($\log T_{\ion{Fe}{xv}} = 6.32$) and the instrumental broadening.

The profiles, both in the examples and in the data at large, are fit surprisingly well by the constrained-width kappa model (Fig.~\ref{fig:kappa_cw_ABCD}).  The fits to A and D, in particular, capture both the peak and the tails, as well as the rest of the profile.  In comparison with the 5-parameter kappa model, the errors are higher (designated $\chi_r^2(\kappa^*)$ and $\bar{\epsilon}(\kappa^*)$ in Tables \ref{tbl:reduced_chisq} and \ref{tbl:unbiased_RSS}).  In addition, the constrained-width fits tend to produce lower values of $\kappa$ to compensate for the width (Fig.~\ref{fig:kappa_cw_histograms}), which is weakly dependent on $\kappa$.  This compensation, however, can only go so far because lower values of $\kappa$ tend to narrow the peak as they widen the tails and have little effect on the width at half-maximum, for a given amplitude.  Profiles A and D show clearly that the constrained-width kappa model can fit the entire length of the profile with minute residuals, even though the width is not allowed to vary.  This is a good indication that the shape of the data is far from Gaussian.

Fig.~\ref{fig:kappa_cw_histograms} shows histograms of the best-fit $\kappa$ and reduced $\chi^2$ values of the constrained-width model.  Like the example profiles, the distribution of $\kappa$ tends to be shifted to lower $\kappa$, but it is much more concentrated on the $\kappa$ scale.  From the distribution of errors (inset), we see again that the profiles can be fit well by constrained-width kappa functions. 


\section{Discussion} \label{sec:discussion}

We applied a smoothness criterion $R_s$ to each profile in the data to determine if it was suitable for fitting; then for profiles whose $R_s>1$, we applied kappa, SG, and DG fits to the data.  Many of the pixels in the first raster were found to be too jagged and therefore eliminated by this smoothness criterion, especially from regions of low intensity and thus poor photon statistics.  The second raster was cleaner, probably due to improved signal-to-noise ratios across the profile.

For the SG model, the least-squares fitting algorithm tried to find a compromise between the peak residual and tail residual, generally favoring the one with lower measurement error.  Given the rigid nature of the Gaussian shape and the non-Gaussianity of the data, no parameters for a SG profile could be found to allow the best-fit curve to capture the peak and the tails simultaneously.

In general, both the averaged unbiased residuals $\bar{\epsilon}$ and the studentized residuals $\chi^2$ were both lower for kappa and DG fits than for the SG.  For high values of $\kappa$, however, the kappa fit was virtually indistinguishable from the SG.  In Profile C, the two fits are practically overlain, with the only visible differences occurring in the continuum.  For this profile, the kappa fit carried a lower $\chi^2$ error than the SG fit but a higher value of $\bar{\epsilon}$.  The primary reason that $\bar{\epsilon}$ can be higher for the kappa model, even though it is a superset of the Gaussian function, is that the optimization routine seeks to minimize the weighted RSS, rather than the absolute residuals.  A secondary reason could be the optimization routine was stuck at a local minimum, due to the failure of the GA algorithm to pass the optimal guess to the LM algorithm.

Although the example profiles presented in this paper are ``representative,'' in the sense that they are not outliers or exceptional cases, our selection could be slightly misleading because the four computed values of $\kappa$ in Figs.~\ref{fig:kappa_SG_weighted_unweighted} and \ref{fig:kappa_SG_BCD} do not occur in equal numbers or probability.  As the histograms of $\kappa$ reveal, the occurrence of high values of $\kappa$ close to Maxwellian/Gaussian only occur in areas of low intensity and could be an artifact of poor photon statistics.  In the 5-parameter kappa model, almost all the profiles with moderate to high intensity tend to have $\kappa<10$, and in the constrained-width kappa model, $\kappa<5$.




While fitting to a complex profile, such as the DG, is feasible, the robustness of the solution delivered by the fit algorithm suffers with each additional component.  In other words, the likelihood of encountering the global minimum RSS is reduced by the number of degrees of freedom.  While some profiles are clearly bi- or multimodal and must be fit as such, most of the profiles encountered for the \ion{Fe}{xv} line at 284.16~\AA\ clearly have a single peak.

The DG model is unique among the three presented here in that it allows for profile asymmetry.  One alternative solution to allow for slight asymmetry in the kappa and SG models would be to allow for a non-constant background, such as a sloped line.  Asymmetry of profiles is an important issue because it is a common feature of spectral data.  However, the fact that EIS profiles, excluding continuum background emission, are essentially represented by 6-12 data points could preclude fitting to such complex profiles as DG, which requires a minimum of 7 degrees of freedom.  A major problem for all three models is that they severely overfit the data, due to poor sampling.  This problem would be resolved by improvement in spectral resolution in future missions.

Given the current spectral resolution, however, reducing the number of free parameters is one way to avoid overfitting the data.  By constraining the width to be a function only of instrumental broadening and equivalent thermal velocity (and therefore dependent on $\kappa$), we are able to keep the same number of free parameters in the constrained-width kappa model as in the SG model, namely 4, including a flat background.  The results of the constrained-width kappa fits reveal that for the majority of profiles in the data, no microturbulence is needed to account for excess widths, as predicted by \citet{Scudder92b}.




While our results pertain to ion populations, \citet{Feldman:2007p16223} have taken a different approach to testing for the presence of suprathermal {\em electron} distributions.  Using {\em SOHO}/SUMER \citep{Wilhelm:1995p14771}, they analyze the intensities of lines from three coronal He-like ions: \ion{Ne}{ix}, \ion{Mg}{xi} and \ion{Si}{xiii} and find no evidence of non-thermal electrons, pointing to the absence of emission from \ion{Si}{xiii} as a constraint on the electron energy tail.  Another study which considers non-thermal electrons is \citet{Muglach:2010p16750} combine SUMER and EIS data and explore the possibility of modeling the transition region emission spectrum using a non-thermal electron distribution consisting of the combination of two Maxwellians \citep[cf.][]{MakPierRil97}.  Although this model does not satisfactorily explain the observed spectral line intensities, their model does not discount the possibility of other descriptions of non-thermal populations, such as the kappa distribution.

A major point of contention that may arise, regarding kappa distributions of coronal ions, is that, while electrons may be collisionless enough to escape thermalization, ions such as \ion{Fe}{xv} carry too high a charge to achieve the same.  According to \citet{Spitzer}, the collision frequency of an ion population should increase as $Z^4$, where $Z$ is the charge state of the ion.  We compute the classical Spitzer collision times in Appendix \ref{app:timescales}, as well as the ionization and recombination times for \ion{Fe}{xv}.  By Spitzer's formula, we find the collision time of \ion{Fe}{xv} to be indeed shorter than its lifetime, but only by a factor of about 2.  Nevertheless, \citet{Scudder92b} argues that the Spitzer formula grossly overestimates the collision frequency, being based on a first-order linearization that assumes small test-particle velocities and Chapman-Enskog theory.  However, as noted in \S \ref{sec:intro_kappa}, collisionality is velocity-dependent in non-thermal rarefied plasmas, and electric fields in plasmas such as the solar transition region and corona can help to maintain high-energy, nearly collisionless, tails of particle distributions \citep{RouDup80a}.  In addition, \citet{PierLamy03} and \citet{PierLamyLem04} argue that kappa distributions of ions in the corona, with low values of $\kappa$, are necessary to explain measured bulk velocities in the solar wind. 

We further note that, in his analysis of the data of \citet{Nicolas82}, \citet{Scudder92b} conjectured that ions in the transition region and corona should have an average $\kappa$ value of about 2.2, which is in good agreement with the results of both kappa model fits (Figs.~\ref{fig:kappa_histograms} and \ref{fig:kappa_cw_histograms}).

\section{Acknowledgments}

The authors wish to thank H.~Peter, E.~Dzif{\v c}\'akov\'a, J.~Dudik, and J.~Scudder for sharing their insights.  The present work was supported by the Onderzoekfonds KU Leuven (Research Fund KU Leuven) and by the European Commission’s Seventh Framework Programme (FP7/2007-2013) under the Grant Agreement No. 218816 (SOTERIA project, www. soteria-space.eu) and No. 263340 (SWIFF project, www.swiff.eu).  DRW acknowledges a UCL University Fellowship and STFC Rolling Grant ST/H00260X/1 for the funding of this research.

{\small
\bibliographystyle{aa}
\bibliography{myrefs}
}
\begin{appendix}

\section{Profile smoothness} \label{app:smoothness}

For a theoretically continuous emission profile $u(x)$, $x \in \Omega \subset \mathbb{R}$, we could describe the functional properties of $u(x)$ in terms of various Lebesgue norms.  We recall the standard $L^p$ norm
\[
\Vert u \Vert_p = \left\{ \int_\Omega u(x)^p dx \right\}^{1/p} \ .
\]

In a Sobolev space $W^{k,p}$, we could define a seminorm
\begin{equation}
\vert u \vert_{j,2} = \left\{ \int_\Omega \left| D^j u(x) \right|^2 dx \right\}^{1/2} \ ,
\end{equation}
\noindent which would describe the total ``power'' of the $j$th derivative of $u(x)$, i.e., $u''(x)$.  Such a measure would not a complete norm for the Sobolev space $W^{j,2}$ and is therefore a seminorm.  It is worth noting that Sobolev seminorms $\vert u \vert_{j,2}$ and $\vert u \vert_{k,2}$ would have distinct physical dimensions for $j \neq k$.

Sobolev seminorms can be used to describe the smoothness of functions in $\mathcal{C}^j(\Omega)$, that is, functions that are continuous and $j$-times differentiable.  While emission profiles can in theory appear continuous, there is no guarantee of Lipschitz continuity and even less of their being differentiable.  Furthermore, in practice, it is sampled as a set of photon measurements at discrete wavelength values, with appreciable measurement error.  However, in lieu of actual derivatives, we can still compute difference functions (i.e., finite differences) which are meaningful at least up to first and second order.

Therefore, we define discrete forms of the Sobolev seminorms,
\begin{eqnarray}
\left| I(\lambda) \right|_{1,2} &=& \left\{ \sum_{i=1}^{N-1} \left| \frac{I(\lambda_{i+1}) - I(\lambda_i)}{\Delta \lambda} \right|^2 \right\}^{1/2} \\
\left| I(\lambda) \right|_{2,2} &=& \left\{ \sum_{i=2}^{N-1} \left| \frac{I(\lambda_{i+1}) - 2 I(\lambda_i) + I(\lambda_{i-1})}{(\Delta \lambda)^2} \right|^2 \right\}^{1/2}
\end{eqnarray}
\noindent To take the ratio between the two seminorms requires dimensional and renormalization factors; thus we define our smoothness measure
\begin{equation}
R_s \left(I(\lambda)\right) = \left(\frac{N-2}{N-1}\right) \frac{\left| I(\lambda) \right|_{1,2}}{\left| I(\lambda) \right|_{2,2} \Delta \lambda}
\end{equation}

\section{\ion{Fe}{xv} time scales in the solar corona.} \label{app:timescales}

According to \cite{Spitzer}, the self-collision time $t_c$ is given by
\begin{equation}
t_c = \frac{11.4 \mu^{\frac{1}{2}}T^{\frac{3}{2}}}{nZ^4\,\ln\Lambda}
\label{tself}
\end{equation}
where $\mu = 55.44$ for iron, $T = 2\times 10^6$~K, the peak formation temperature of \ion{Fe}{xv}, and $Z = 14$, and the Coulomb logarithm is given by
\begin{equation}
\ln\Lambda = \ln\left[\frac{3}{2Z^2e^3} \left(\frac{k^3T^3}{\pi n_e}\right)^\frac{1}{2}\right] = 56.83 .
\label{coulomb}
\end{equation}
We use the the coronal abundance of iron, $A_{\mathrm{Fe}}$, given by
\cite{Feldman:1992p16576} and the peak ionisation fraction of \ion{Fe}{xv}, $f$, given by \cite{Bryans09} to derive the number density of ions, $n$:
\begin{equation}
n = \frac{n_H}{n_e}\,A_{\mathrm{Fe}}\,n_e\,f = 1.96 \times 10^4
\label{iondens}
\end{equation}
with typical values for the proton/electron ratio, $n_H / n_e =  0.8$, and electron density, $n_e = 10^9$~cm$^{-3}$. Combining Eqs.~\ref{tself}, \ref{coulomb} and \ref{iondens} then gives a self-collision time, $t_c = 5.6$~s.




We can  compare these with the ionisation and recombination times to and from \ion{Fe}{xv} calculated by \cite{Golub:1989p15607}:
\begin{itemize}
\item $t_{Z-1 \rightarrow Z}^{ion}$ = 9.0~s
\item $t_{Z \rightarrow Z+1}^{ion}$ = 16~s
\item $t_{Z+1 \rightarrow Z}^{rec}$ = 9.9~s
\item $t_{Z \rightarrow Z-1}^{rec}$ = 7.0~s .
\end{itemize}
The self-collision time calculated using \citeauthor{Spitzer} is shorter than these timescales, but see Section~\ref{sec:discussion} for a discussion of this in the context of \cite{Scudder92b}.

\end{appendix}

\end{document}